\documentclass[12pt]{iopart}

\usepackage{iopams,cite}
\usepackage{graphicx}
\newcommand*\vs{\boldsymbol}
\begin{document}
\title{First integrals of the axisymmetric shape equation of lipid membranes}

\author{Y H Zhang$^1$, Z McDargh$^2$, Z C Tu$^1$,}

\address{$^1$ Department of Physics, Beijing Normal University, Beijing 100875, China}
\address{$^2$ Carnegie Mellon University, 5000 Forbes Avenue, Pittsburgh PA 15213, USA}
\ead{tuzc@bnu.edu.cn}
\vspace{10pt}

\begin{abstract}
The shape equation of lipid membranes~[Zhong-can and Helfrich(1987) PRL 59 2486] is a fourth-order partial differential equation. Under the axisymmetric condition, this equation was transformed into a second-order ordinary differential equation (ODE) by Zheng and Liu~[Zheng and Liu(1993) PRE 48 2856]. Here we try to further reduce this second-order ODE to a first-order ODE by seeking for first integrals according to the Noether theorem. We indeed find a first integral when the second-order ODE is invariant under conformation transformations. We obtain the mechanical interpretation of the first integral by using the membrane stress tensor.
\end{abstract}

\pacs{00.02, 60.68, 80.87}
\vspace{2pc}
\noindent{\it Keywords}: lipid membrane, shape equation, first integral, Noether theorem

%\submitto{\jpa}
%\maketitle

\section{Introduction}
The elasticity of membranes and shells has drawn much research attention. In 1812, Poisson~\cite{Poisson1833} proposed a functional for the bending energy of a shell
\begin{eqnarray}
F=\frac{k_c}{2}\int_MH^2\rmd A,\label{eq55}
\end{eqnarray}
where $H$ represents the mean curvature of the shell surface, $M$ and $\rmd A$ represent the shell surface and its area element, respectively. Equation (\ref{eq55}) is known as the Willmore functional in mathematics, and is invariant under conformal transformations of the embedding space. The equilibrium configurations of the shell minimize the Willmore functional, and must satisfy the Willmore equation
\begin{eqnarray}
\nabla^2H+2H(H^2-K)=0,\label{eq56}
\end{eqnarray}
where $K$ represents the Gauss curvature of the shell surface, and $\nabla^2$ is the Laplace-Beltrami operator on a 2-dimensional surface. For compact surfaces in 3-dimensional Euclidian space, Willmore proved that round spheres and their images under conformal transformations correspond to the least minimum of the Willmore functional. Thus the Willmore functional takes values no less than $4\pi$ for all compact surfaces. He further conjectured that the values of the Willmore functional are no less than $2\pi^2$ for compact surfaces of genus one in 3-dimensional Euclidian space. With this topology, the Willmore tori and their images under conformal transformations are the least minimum. The Willmore tori are special tori with the ratio of their two generating radii being $\sqrt2$. The Willmore conjecture was recently proved by Marques and Neves via min-max theory~\cite{Marques2014}. Willmore surfaces can also arise from the mKdV equation with the deformation technique. Tek investigated the Weingarten and Willmore-like surfaces arising from the spectral deformation of the Lax pair for the mKdV equation~\cite{Tek2007}.

In 1973, Helfrich proposed the spontaneous curvature model of lipid membranes~\cite{Helfrich1973}. He supposed that a lipid membrane could be treated as a liquid crystalline thin film. Based on the elastic theory of liquid crystals, Helfrich derived the bending energy of the membrane
\begin{eqnarray}
F_\mathrm{H}=\int_M\left[(k_c/2)(2H+c_0)^2+\bar{k}K\right]\rmd A,\label{eq57}
\end{eqnarray}
where $k_c>0$ and $\bar{k}$ are bending moduli of the membrane, and $c_0$ is the spontaneous curvature reflecting the difference of environmental factors or chemical constituents between the two leaflets of the membrane. The Helfrich functional prompted a flourishing theory of lipid membrane elasticity~\cite{ZC1987,ZC1989,Konopelchenko1997,Lipowsky1991,Landolfi2003,Deuling1976,Seifert1997,Zhong-Can1999b,Guven2002,Guven2003,Guven2004a,Guven2005b,Guven2005a,Tu2004,Tu2013}.

Since the area of lipid bilayer is almost inextensible and the volume enclosed by a lipid vesicle is hardly compressed, the shape of the lipid vesicle in equilibrium corresponds to the minimal value of (\ref{eq57}) at fixed volume and area. We therefore introduce two Lagrange multipliers to enforce these constraints, and then minimize the following extended functional
\begin{eqnarray}
F=\int_M\left[(k_c/2)(2H+c_0)^2+\bar{k}K+\lambda\right]\rmd A+p\int\rmd V,\label{eq58}
\end{eqnarray}
where the Lagrange multipliers $\lambda$ and $p$ can be interpreted as the surface tension and osmotic pressure, respectively. The first order variation of (\ref{eq58}) leads to the shape equation of lipid membranes~\cite{ZC1987,ZC1989}
\begin{eqnarray}
2k_c\nabla^2H+k_c(2H+c_0)(2H^2-2K-c_0H)-2\lambda H+p=0.\label{eq59}
\end{eqnarray}
The above partial differential equation (PDE) degenerates into the Willmore equation when $p$, $\lambda$, and $c_0$ are vanishing. Konopelchenko investigated (\ref{eq58}) and (\ref{eq59}) by introducing the so-called generalized Weierstrass representation~\cite{Konopelchenko1997}. He found that it could be reduced to several solvable equations, such as the Liouville equation or the Schr\"{o}dinger equation. Following this framework, Landolfi provided a new type of periodic developable surfaces and further discussed the statistical properties of membranes described by the Helfrich energy~\cite{Landolfi2003}. Numerical study reveals that the biconcave discoid configuration indeed corresponds to the minimum value of (\ref{eq58}) under appropriate physiological conditions~\cite{Deuling1976}, which may explain the biconcave shape of normal red blood cells.

\begin{figure}[h]\label{pic1}
\includegraphics[width=3in]{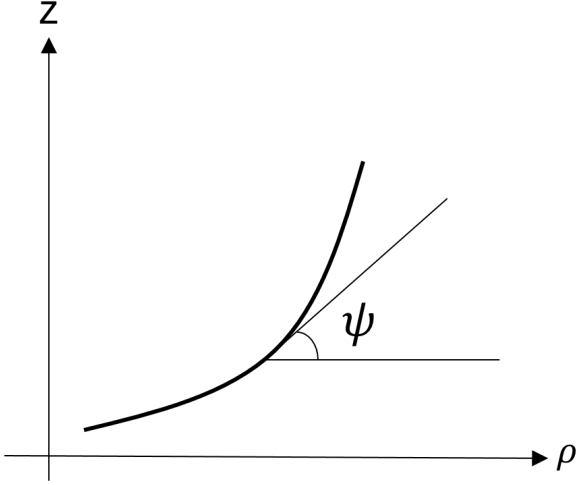}
\caption{Schematic of the profile curve.}
\end{figure}

One can generate an axisymmetric surface by rotating a planar curve $z=z(\rho)$, with $\rho$ being the radius of revolution around the $z$-axis. The axisymmetric surface may be expressed as:
\begin{eqnarray}
x=\rho\cos{\phi},\quad y=\rho\sin{\phi},\quad z=\int\tan{\psi(\rho)}\rmd\rho,\label{eq60}
\end{eqnarray}
where $\phi$ is the azimuthal angle and $\psi$ is the tangent angle of the profile curve as shown in Figure \ref{pic1}.

For axisymmetric surfaces, the shape equation (\ref{eq59}) can of course be written as an ordinary differential equation (ODE) rather than a PDE, making it much easier to solve. Using the parametrization in equation (\ref{eq60}), it becomes \cite{Hu1993}
\begin{eqnarray}
\fl-\frac{\cos{\psi}}{\rho}\left\{\rho\cos{\psi}\left[\frac{(\rho\sin{\psi})'}{\rho}\right]'\right\}'
-\frac12\left[\frac{(\rho\sin{\psi})'}{\rho}\right]^3\nonumber\\+\frac{(\rho\sin{\psi})'(\sin^2{\psi})'}{\rho^2}
-\frac{c_0(\sin^2{\psi})'}{\rho}+\frac{\tilde{\lambda}(\rho\sin{\psi})'}{\rho}+\tilde{p}=0,\label{eq61}
\end{eqnarray}
where $\tilde{\lambda}\equiv\lambda/k_c+c_0^2/2$, $\tilde{p}\equiv p/k_c$, $\psi'=\rmd\psi/\rmd\rho$, $\psi''=\rmd^2\psi/\rmd\rho^2$. A first integral of (\ref{eq61}) was found in~\cite{Zheng1993}, and the above third-order ODE was reduced to the following second-order ODE:
\begin{eqnarray}
\fl \rho\cos^3\psi\psi''-\frac{\rho}{2}\cos^2\psi\sin\psi\psi'^2+\cos^3\psi\psi'\nonumber\\
+\frac{\sin^3\psi}{2\rho}+c_0\sin^2\psi-\frac{\tilde{\lambda}\rho^2+1}{\rho}\sin\psi
-\frac{\tilde{p}}{2}\rho^2+\Omega_0=0. \label{eq5}
\end{eqnarray}
where $\Omega_0$ is the first integral of (\ref{eq61}). This constant can be interpreted as tension in the membrane along the symmetry axis, and is a result of the axisymmetry of the surface. Researchers have found several special solutions of (\ref{eq5}), including minimal surfaces (catenoid, helicoid, \emph{etc.}), constant mean curvature surfaces (sphere, cylinder, unduloid~\cite{Naito1995,Mladenov2002}, \emph{etc.}), Willmore surfaces (Clifford torus~\cite{Ou-Yang1990}, Dupin Cyclide~\cite{Ou-Yang1993}, inverted catenoid~\cite{Castro2007}, \emph{etc.}), cylinder-like surfaces~\cite{Ou-Yang1996,Vassilev2008,ZhouXiao-Hua2010}, and circular biconcave discoid~\cite{Naito1993,Naito1996}.

 As (\ref{eq5}) is a second-order ODE, a natural question is whether we can further transform it into a first-order ODE by constructing another first integral of (\ref{eq5}), which is the main goal of this paper. Usually, a second-order ODE may be regarded as the equation of motion of a mechanical system in mathematics. Thus, we might be able to transform~(\ref{eq5}) into a first-order ODE by finding first integrals for a given mechanical system. In a previous work by Capovilla \emph{et al.}.~\cite{Guven2005b,Guven2005a}, they treated the Helfrich energy (\ref{eq57}) as a kind of action describing the motion of a particle, and gave the shape equation by solving the corresponding Hamilton equations. Then (\ref{eq5}) could emerge naturally from their framework with axial symmetry. Here we start from (\ref{eq5}) directly, and construct the corresponding Lagrangian by solving the inverse problem of the calculus of variations. Our aim is to find first integrals of (\ref{eq5}). The rest of the paper is organized as follows: in section 2, we will construct the Lagrangian corresponding to shape equation~(\ref{eq5}). In section 3, we will briefly introduce the generalized Noether theorem, which we will use to look for first integrals of shape equation (\ref{eq5}). In the special case of vanishing $p$, $c_0$, $\lambda$ and $\Omega_0$, we successfully achieve a first integral and the corresponding special solution to the shape equation. In section 4, we find that this special solution can also be derived from the Hamilton-Jacobi equation. Section 5 generalizes the first integral to cases with $\Omega_0\ne 0$ and demonstrates that we can also connect the origins of the first integral and its mechanical interpretation with the membrane's stress tensor. Section 6 contains a summary and discussion of our results and outlook.

\section{Lagrangian corresponding to the shape equation}

In this section, we will construct the Lagrangian corresponding to (\ref{eq5}).

Assume that we have a Lagrangian with the following very general form:
\begin{eqnarray}
  L=\frac12\alpha(t,u) \dot{u}^2+\beta(t,u) \dot{u}+\gamma(t,u). \label{eq20}
\end{eqnarray}
and an arbitrary equation of motion:
\begin{eqnarray}
  a(t,u)\ddot{u}+b(t,u)\dot{u}^2+c(t,u)\dot{u}+d(t,u)=0,\label{eq22}
\end{eqnarray}
where $\ddot{u}=\rmd^2u/\rmd t^2$, $\dot{u}=\rmd u/\rmd t$. The Euler-Lagrange equation $\frac{\partial L}{\partial u}-\frac{\rmd}{\rmd t}\left(\frac{\partial L}{\partial\dot{u}}\right)=0$ corresponding to (\ref{eq20}) may be expressed as:
\begin{eqnarray}
  \alpha \ddot{u}+\frac12\frac{\partial\alpha}{\partial u}\dot{u}^2+\frac{\partial\alpha}{\partial t}\dot{u}+\frac{\partial\beta}{\partial t}-\frac{\partial\gamma}{\partial u}=0.\label{eq21}
\end{eqnarray}
Equations (\ref{eq22}) and (\ref{eq21}) have the same solution if there is a non-vanishing function $f$ such that
\begin{equation}
 \eqalign{
  \frac{\alpha}{a}=\frac{\frac12\frac{\partial\alpha}{\partial u}}{b}=\frac{\frac{\partial\alpha}{\partial t}}{c}=\frac{\frac{\partial\beta}{\partial t}-\frac{\partial\gamma}{\partial u}}{d}=f.\label{eq23}}
\end{equation}

Now compare the shape equation (\ref{eq5}) with equation of motion (\ref{eq22}), by identifying $t$ and $u$ with $\rho$ and $\psi$, respectively. Then from (\ref{eq23}) we derive:
\begin{eqnarray}
  \eqalign{
  \alpha=\rho\cos\psi,\quad f=\sec^2\psi,\cr
  \frac{\partial\beta}{\partial\rho}-\frac{\partial\gamma}{\partial\psi}=\sec^2\psi
  \left(\frac{\sin^3\psi}{2\rho}+c_0\sin^2\psi-\frac{\tilde{\lambda}\rho^2+1}{\rho}\sin\psi
  -\frac{\tilde{p}}{2}\rho^2+\Omega_0\right).\label{eq24}}
\end{eqnarray}

The functions $\beta$ and $\gamma$ contain some degrees of freedom which we are free to choose. This is because Lagrangians that differ by a divergence lead to the same equations of motion. Consider two different Lagrangians of the form of (\ref{eq20}):
\begin{eqnarray}
  L=\frac12\alpha\psi'^2+\beta\psi'+\gamma,\label{eq25}\\
  \tilde{L}=\frac12\alpha\psi'^2+\tilde{\beta}\psi'+\tilde{\gamma}.	\label{eq26}
\end{eqnarray}
Taking the difference between (\ref{eq25}) and (\ref{eq26}),
\begin{eqnarray}
  L-\tilde{L}=\left(\beta-\tilde{\beta}\right)\psi'+\left(\gamma-\tilde{\gamma}\right).\label{eq27}
\end{eqnarray}
On the other hand, we have
\begin{eqnarray}
  \frac{\partial\beta}{\partial\rho}-\frac{\partial\gamma}{\partial\psi}=
  \frac{\partial\tilde{\beta}}{\partial\rho}-\frac{\partial\tilde{\gamma}}{\partial\psi},\label{eq28}
\end{eqnarray}
because equation (\ref{eq24}) requires the quantity $\partial \beta/\partial \rho - \partial \gamma/\partial \psi$ to have a specific form. We can rearrange this to give
\begin{eqnarray}
  \frac{\partial}{\partial\rho}\left(\beta-\tilde{\beta}\right)=
  \frac{\partial}{\partial\psi}\left(\gamma-\tilde{\gamma}\right).	\label{eq29}
\end{eqnarray}
This equation implies that there must exist a function $\Phi(\rho,\psi)$ satisfying
\begin{eqnarray}
  \frac{\partial\Phi}{\partial\psi}=\beta-\tilde{\beta},\quad\frac{\partial\Phi}{\partial\rho}=\gamma-\tilde{\gamma},\label{eq7}
\end{eqnarray}
which leads to
\begin{eqnarray}
  L-\tilde{L}=\frac{\rmd}{\rmd\rho}\Phi(\rho,\psi).\label{eq30}
\end{eqnarray}

 Thus the difference between (\ref{eq25}) and (\ref{eq26}) is just a total derivative of some function, and the equation of motion corresponding to both equations are therefore the same. For the sake of simplicity, we take $\beta=0$, and the Lagrangian is expressed as
 \begin{equation}
   L=\frac{\rho\cos\psi}{2}\psi'^2+\frac{\tan{\psi}\sin{\psi}}{2\rho}+c_0\psi+\tilde{\lambda}\rho\sec{\psi}-\left(c_0+\Omega_0-\frac{\tilde{p}}{2}\rho^2\right)\tan\psi.~~\label{eq31}
 \end{equation}

\section{Noether theorem and conservation law}

In Newtonian mechanics, Noether's theorem explains the connection between symmetries and conservation laws. Since we have the Lagrangian of the equation of motion, we can construct the first integrals of the equation of motion by using Noether's theorem and looking for symmetries of this Lagrangian.

\subsection{The generalized Noether theorem}
From the generalized Noether theorem we know that~\cite{Ibragimov1969} if and only if the value of the variational integral $\int L(t,u,\dot{u})\rmd u$ is invariant at the extrema under a continuous transformation group with generator
\begin{eqnarray}
  X=\xi\frac{\partial}{\partial t}+\eta\frac{\partial}{\partial u},\label{eq15}
\end{eqnarray}
then the scalar
\begin{eqnarray}
  I=\xi L+\left(\eta-\xi \dot{u}\right)\frac{\partial L}{\partial \dot{u}}\label{eq16}
\end{eqnarray}
provides a first integral for the Euler-Lagrange equation corresponding to $L$. We can test for this invariance by checking that there exists a function $F$ such that
\begin{eqnarray}
  X\left(L\right)+L\dot{\xi}=F\frac{\delta L}{\delta u},\label{eq17}
\end{eqnarray}
where
\begin{eqnarray}
  \frac{\delta L}{\delta u}\equiv\frac{\partial L}{\partial u}-\frac{\rmd}{\rmd t}\left(\frac{\partial L}{\partial \dot{u}}\right),\quad X\left(L\right)=\xi\frac{\partial L}{\partial t}+\eta\frac{\partial L}{\partial u}+\left(\dot{\eta}-\dot{u}\dot{\xi}\right)\frac{\partial L}{\partial \dot{u}}.\label{eq18}
\end{eqnarray}
We demand that $F\neq\eta-\xi \dot{u}$ because the first integral (\ref{eq16}) is otherwise trivial.

Alternatively, the infinitesimal test condition (\ref{eq17}) could be replaced by the divergence condition~\cite{Ibragimov1969}
\begin{eqnarray}
  X\left(L\right)+L\dot{\xi}=F\frac{\delta L}{\delta u}+\dot{B}. \label{eq34}
\end{eqnarray}
This is equivalent to equation (\ref{eq17}) modulo a total derivative term in the Lagrangian. Using this condition, the corresponding first integrals become
\begin{eqnarray}
  I=\xi L+\left(\eta-\xi \dot{u}\right)\frac{\partial L}{\partial \dot{u}}-B.\label{eq35}
\end{eqnarray}

\subsection{First integral of the axisymmetric shape equation of lipid membranes}

 For the sake of brevity, we rewrite the Lagrangian in (\ref{eq31}) in the form
 \begin{eqnarray}
   L(\rho,\psi,\psi')=\frac{\rho\cos\psi}{2}\psi'^2-V(\rho,\psi),\label{eq32}
 \end{eqnarray}
 where we now consider all the terms not containing $\psi'$ as constituting a potential energy
 \begin{eqnarray}
    V(\rho,\psi)=-\frac{\tan{\psi}\sin{\psi}}{2\rho}-c_0\psi-\tilde{\lambda}\rho\sec{\psi}+\left(c_0+\Omega_0-\frac{\tilde{p}}{2}\rho^2\right)\tan\psi.\label{eq33}
 \end{eqnarray}
 Applying conditions (\ref{eq18}) and (\ref{eq34}) for invariance of the action to (\ref{eq32}) with $u\rightarrow\psi$ and $t\rightarrow\rho$, we find that, if the first integral (\ref{eq35}) exists, then $B(\rho,\psi,\psi')$, $\eta(\rho,\psi,\psi')$, $\xi(\rho,\psi,\psi')$ and $F(\rho,\psi,\psi')$ satisfy the following equation:
\begin{eqnarray}
  \fl \left[2\rho\cos\psi F-2B_{\psi'}-\left(\rho\psi'^2\cos{\psi}+2V\right)\xi_{\psi'}+2\rho\psi'\cos{\psi}\eta_{\psi'}\right]\psi''\nonumber\\
  -\rho\cos\psi\xi_\psi\psi'^3+\left[\cos\psi\left(\xi-\rho\xi_\rho+2\rho\eta_\psi\right)-\rho\sin\psi\left(F+\eta\right)\right]\psi'^2\nonumber\\
  +2\left[\cos\psi\left(F+\rho\eta_\rho\right)-B_\psi-V\xi_\psi\right]\psi'\nonumber\\
  -2\xi V_\rho+2\left(F-\eta\right)V_\psi-2B_\rho-2V\xi_\rho=0.\label{eq36}
\end{eqnarray}
In the above equation, $\psi''$, $\psi'$ and $\psi$ should be regarded as independent variables. The subscripts $\rho$, $\psi$ and $\psi'$ denote the partial derivatives with respect to $\rho$, $\psi$ and $\psi'$. Naturally, we can separate (\ref{eq36}) into two equations according to the order of $\psi''$:
\begin{eqnarray}
\fl2\rho\cos\psi F-2B_{\psi'}-\left(\rho\psi'^2\cos{\psi}+2V\right)\xi_{\psi'}+2\rho\psi'\cos{\psi}\eta_{\psi'}=0,\label{eq37}
\end{eqnarray}
and
\begin{eqnarray}
  \fl-\rho\cos\psi\xi_\psi\psi'^3+\left[-\rho\sin\psi\left(F+\eta\right)+\cos\psi\left(\xi-\rho\xi_\rho+2\rho\eta_\psi\right)\right]\psi'^2\nonumber\\
  +2\left[\cos\psi\left(F+\rho\eta_\rho\right)-B_\psi-V\xi_\psi\right]\psi'\nonumber\\
  -2\xi V_\rho+2\left(F-\eta\right)V_\psi-2B_\rho-2V\xi_\rho=0.\label{eq38}
\end{eqnarray}

We have investigated the situation that $B$, $\eta$, $\xi$ and $F$ do not explicitly contain $\psi'$, and find that the solutions to (\ref{eq37}) and (\ref{eq38}) only exist if the parameters $c_0$, $\tilde{\lambda}$, $\tilde{p}$ and $\Omega_0$ all vanish. By taking these parameters to vanish, (\ref{eq5}) in fact becomes the axisymmetric Willmore equation. The details of these calculations will be shown in the appendix. The final result is
\begin{eqnarray}
B=0,\quad\eta=0,\quad\xi=2\rho.\label{eq39}
\end{eqnarray}
Thus $X$ is a generator of dilations, a subset of the well-known corformal invariance of the Willmore functional. By substituting (\ref{eq39}) into (\ref{eq35}), we obtain its first integral
\begin{eqnarray}
  I=\rho^2\cos\psi\psi'^2-\frac{\sin^2\psi}{\cos\psi} = \rho^2\sec \psi \left (\cos^2 \psi \psi'^2-\frac{\sin^2\psi}{\rho^2} \right). \label{eq12}
\end{eqnarray}
Note that $\sin\psi/\rho$ and $-\psi' \cos \psi$ are the principal curvatures of the membrane surface.
The solution of the ODE (\ref{eq12}) can then be obtained by the quadrature
 \begin{eqnarray}
  \rho=\rho_0\mathrm{exp}\left(\pm\int\frac{\cos{\psi}}{\sqrt{I\cos{\psi}+\sin^2{\psi}}}\rmd\psi\right),\label{eq2}
\end{eqnarray}
with $\rho_0$ being an unknown constant.

In fact, $I$ is the only independent parameter in this solution. The profile curve of the surface corresponding to different $I$ is depicted in Figure 2. Choosing the negative sign in the exponent in equation (\ref{eq2}) leads to solutions with $\rho/\rho_0 \le 1$, while choosing the positive sign leads to solutions with $\rho/\rho_0 \ge 1$. For example, if we take $I=0$, we find the two solutions, $\rho=\rho_0\sin\psi$ or $\rho=\rho_0\csc\psi$, depending on which branch we choose for the square root in the exponent. The solution $\rho=\rho_0\sin\psi$ describes a sphere, and is shown as the the solid with $\rho/\rho_0 \le 1$. The solution $\rho=\rho_0\csc\psi$ describes a catenoid, and is shown as the solid line with $\rho/\rho_0 \ge 1$. Solutions with $I\neq0$ usually correspond to non-closed configurations.  The result is similar when $I$ is negative, with switching the signs in (\ref{eq2}).

\begin{figure}[h]\label{pic2}
\includegraphics[width=4.1in]{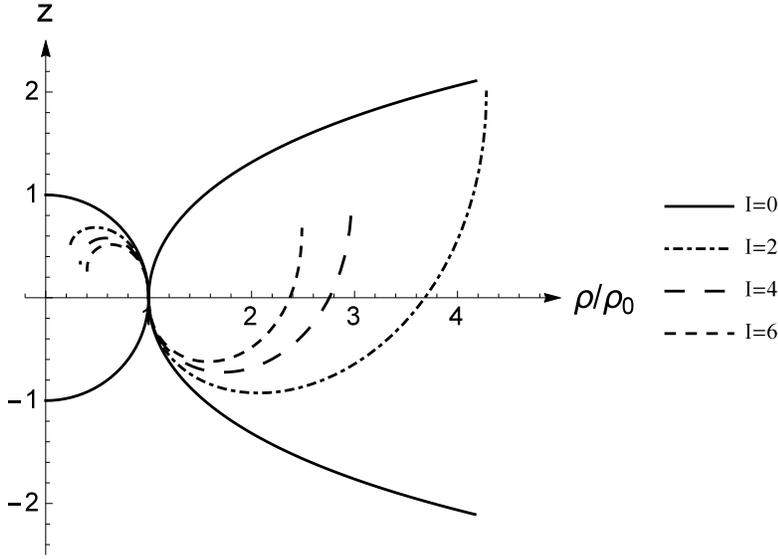}
\caption{Profile curves with different $I$}
\end{figure}

We can also rewrite equation (\ref{eq12}) as
\begin{eqnarray}
  \rho\cos\psi\psi'\pm\sqrt{I\cos\psi+\sin^2\psi}=0.\label{eq13}
\end{eqnarray}
Using the fact that $\tan\psi=\rmd z/\rmd\rho$, this can in turn be transformed into
\begin{eqnarray}
  \frac{\rmd^2z}{\rmd\rho^2}=\pm\frac1\rho\left[\left(\frac{\rmd z}{\rmd\rho}\right)^2+1\right]\sqrt{\left(\frac{\rmd z}{\rmd\rho}\right)^2+I\sqrt{\left(\frac{\rmd z}{\rmd\rho}\right)^2+1}}.\label{eq4}
\end{eqnarray}
This form of Willmore equation was studied in~\cite{vassilev2004,Vassilev2014} through the geometric (point) Lie symmetry group analysis.

\section{Hamilton-Jacobi equation}
For the case of vanishing $c_0$, $\tilde{\lambda}$, $\tilde{p}$ and $\Omega_0$, our first integral can also be obtained via the Hamilton-Jacobi equation.

As we have constructed the Lagrangian (\ref{eq32}), the conjugate momentum to the coordinate $\psi$ is
\begin{eqnarray}
p\equiv\frac{\partial L}{\partial \psi'}=\rho\cos{\psi}\psi'.\label{eq41}
\end{eqnarray}
We further obtain the Hamiltonian
\begin{eqnarray}
H=p\psi'-L=\frac{p^2}{2\rho\cos{\psi}}+V(\rho,\psi).\label{eq42}
\end{eqnarray}
This leads to the corresponding Hamilton-Jacobi equation
\begin{eqnarray}
\frac{\partial S}{\partial\rho}+\frac{1}{2\rho\cos{\psi}}\left(\frac{\partial S}{\partial \psi}\right)^2+V=0,\label{eq43}
\end{eqnarray}
where $S=S(\rho,\psi)$ is the principal function.

When $c_0$, $\tilde{\lambda}$, $\tilde{p}$, $\Omega_0$ vanish, (\ref{eq43}) becomes
\begin{eqnarray}
\frac{\partial S}{\partial\rho}+\frac{1}{2\rho\cos{\psi}}\left(\frac{\partial S}{\partial \psi}\right)^2-\frac{\tan\psi\sin\psi}{2\rho}=0.\label{eq44}
\end{eqnarray}
This equation may be solved by separation of variables. The solution is given by
\begin{eqnarray}
S=\pm\int\sqrt{I'\cos{\psi}+\sin^2{\psi}}\rmd\psi-\frac{I'}{2}\ln{\rho},\label{eq45}
\end{eqnarray}
where $I'$ is a constant.

This method also leads us quickly back to the quadrature for $\psi(\rho)$. The conjugate variable to $I'$ is given by
\begin{eqnarray}
J=\frac{\partial S}{\partial I'}=\frac12\left(\pm\int\frac{\cos{\psi}}{\sqrt{I'\cos{\psi}+\sin^2{\psi}}}\rmd\psi-\ln{\rho}\right).\label{eq46}
\end{eqnarray}
Since $J$ is another constant, its total differential vanishes. Taking advantage of this fact, we obtain
\begin{eqnarray}
2\rmd J = 0=\pm\frac{\cos{\psi}}{\sqrt{I'\cos{\psi}+\sin^2{\psi}}}\rmd\psi-\frac{\rmd\rho}{\rho}.\label{eq6}
\end{eqnarray}
Identifying $I'=I$, we see that agrees with equation (\ref{eq13}). Thus we may derive the first integral (\ref{eq12}) by solving (\ref{eq6}) for $I'$.
%%%%%%%%%%%%%%%%%%%%%%%%%%%%%%%%%%%%%%%%%%%%%%%%%%%%%%%%%%%%%%%%%%%%%%%%%%%%%%%%
\section{Mechanical interpretation}
We can use the membrane's stress tensor $\vs{f}^a$ to identify the origins of our first integral as well as its mechanical interpretation. Furthermore, this method gives a coordinate independent expression for the first integral $I$, and generalizes it to cases when $\Omega_0\ne0$

Generally, the stress tensor of a membrane is given by
\begin{eqnarray}
    \vs{f}^a &= f^{a b} \vs{e}_b + f^a \vs{n},~\textrm{where} \label{decomp} \\
    f^a &= -2 k_c \nabla^a H, \\
    f^{a b} &= k_c K^{ab}(2 H + c_0) - g^{ab} \left[ \frac{k_c}{2} (2 H + c_0)^2+\lambda \right], \label{tan}
\end{eqnarray}
where $K_{a b}$ is the curvature tensor, and the $\vs{e}_a$ are the surface tangent vectors \cite{Guven2002}. In equilibrium, the divergence of the stress tensor balances the local pressure difference across the membrane,
\begin{equation}
    \nabla_a \vs{f}^a = p \vs{n},
\end{equation}
where $\vs{n}$ is the surface normal vector.

As mentioned above, the Willmore functional is invariant under conformal transformations of the ambient space. Consider the subset of conformal transformations known as dilations, i.e. scale transformations. Following ref. \cite{mueller2005}, we can find the Noether current associated with these transformations by considering the energy along with a system of Lagrange multipliers,
\begin{eqnarray}
    E = \int_M \rmd A &\left\{ \mathcal E(g_{ab},K_{ab}) - \vs{f}^a \cdot (\partial_a \vs{X} - \vs{e}_a )
    + \lambda^a_\perp \vs{e}_a \cdot \vs{n} + \lambda_n(\vs{n}^2 -1) \nonumber \right. \\
    &+ \left. \Lambda^{ab}(K_{ab} - \vs{e}_a \cdot \partial_b \vs{n}) + \lambda^{ab}(g_{ab} - \vs{e}_a \cdot \vs{e}_b) \right\}
\end{eqnarray}
where $\vs{X}$ is the embedding of the membrane, and $\mathcal E$ is some energy density that depends only on $g_{ab}$ and $K_{ab}$. The Lagrange multiplier terms allow us to vary $\vs{X}$, $\vs{e}_a$, $K_{ab}$, and $g_{ab}$ independently.

Modulo the Euler-Lagrange equations for the surface, the variation of the energy is
\begin{eqnarray}
	\delta E = -\int_M \rmd A \nabla_a (\vs{f}^a \cdot \delta \vs{X} + \Lambda^{ab} \vs{e}_b \cdot \delta \vs{n}).
\end{eqnarray}
Under an infinitesimal scale transformation, the normal vector is unchanged, while the variation of the embedding is $\delta \vs{X} = \epsilon \vs{X}$. We therefore identify $Q^a = \vs{f}^a \cdot \vs{X}$ as the Noether current associated with scale invariance. Indeed, taking the divergence of this current, we find
\begin{equation}
	\nabla_a( \vs{f}^a \cdot \vs{X}) = p \vs{n} \cdot \vs{X} + \vs{f}^a \cdot \nabla_a \vs{X} = p \vs{n} \cdot \vs{X} + \vs{f}^a \cdot \vs{e}_a = p \vs{n} \cdot \vs{X} +  f^a_{~a},
\end{equation}
where we have used equation (\ref{decomp}) to find the final equality. For the Willmore functional $p = \lambda = c_0 = 0$, and we see that both terms in $\nabla_a Q^a$ vanish.
%The change in energy is therfore
%\begin{eqnarray}
%    \delta E &= - \epsilon \int_M \rmd A \nabla_a( \vs{f}^a \cdot \vs{X} )  \nonumber \\
%    &= - \epsilon \int_M \rmd A \vs{f}^a \cdot \nabla_a \vs{X}  \nonumber \\
%    &= - \epsilon \int_M \rmd A \vs{f}^a \cdot \vs{e}_a  \nonumber \\
%    &= - \epsilon \int_M \rmd A  f^a_a,
%\end{eqnarray}
%where we have used equation (\ref{decomp}) between the third and fourth line. This implies that the trace of the tangential part of the stress tensor vanishes\footnote{I'm not positive if the current argument just implies that the integral must vanish. If we can apply the argument to an arbitrary membrane patch, we can conclude $f^a_a$ vanishes everywhere.}. Indeed, comparing with equation (\ref{tan}), we see that the trace indeed vanishes when $\lambda = c_0 = 0$. But note that the trace can also be reformulated as a divergence:
%\begin{eqnarray}
%    \nabla_a (\vs{f}^a \cdot \vs{X}) = (\nabla_a \vs{f}^a) \cdot \vs{X} + \vs{f}^a \cdot \nabla_a \vs{X} = \vs{f}^a \vs{e}_a = f^a_a.
%\end{eqnarray}
%We therefore identify $\vs{f}^a \cdot \vs{X}$ as the desired Noether current.

\begin{figure}
    \begin{center}
    \includegraphics[height = 0.3\textwidth]{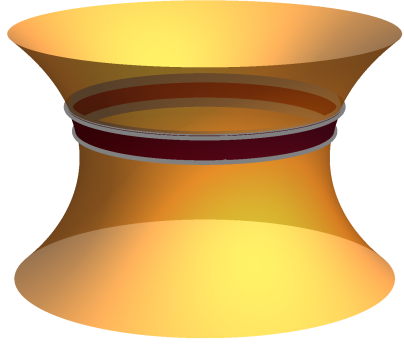}
    \caption{An axisymmetric strip $\Sigma$ (shown in red, with its boundary curves in gray) on a catenoid. \label{patch}}
    \end{center}
\end{figure}

This Noether current leads us to a conservation law for axially symmetric membranes as follows. Integrate $\nabla_a(\vs{f}^a \cdot \vs{X})$ over a strip $\Sigma$ of the surface bounded by two horizontal rings, as shown in figure \ref{patch}.
We can use the divergence theorem to then replace the integral with two line integrals,
\begin{equation}
    \int_\Sigma \rmd A \nabla_a (\vs{f}^a \cdot \vs{X}) =
    \int_{\mathcal C_1} \rmd \phi ~ \rho \vs{f}^a \cdot \vs{X} t_a  - \int_{\mathcal C_2} \rmd \phi ~ \rho \vs{f}^a \cdot \vs{X} t_a = 0,
\end{equation}
where $\mathcal C_1$ and $\mathcal C_2$ are the upper and lower boundary curves of the strip, and $t_a$ is the surface vector tangent to the meridians. Since the patch of membrane was chosen arbitrarily, we conclude that the integral,
\begin{equation}
    \int_{\mathcal C} \rmd \phi ~ \rho \vs{f}^a \cdot \vs{X} t_a = 2\pi \rho \vs{f}^a \cdot \vs{X} t_a
\end{equation}
is constant for all rings $\mathcal C$.

Inserting our parametrization of the surface $\vs{X} = ( \rho \cos \phi, \rho \sin \phi, z)^T$, we find
\begin{eqnarray}
    2\pi \rho \vs{f}^a \cdot \vs{X} t_a &= \pi  k_c \left(\rho^2 \psi'^2 \cos \psi- \tan \psi (\sin \psi-2 \rho  \Omega_0 )-2 \Omega_0  z\right) \\
    &= \pi k_c I +2 \pi k_c \Omega_0( \rho \tan \psi - z),
\end{eqnarray}
where we have used the first integral in equation (\ref{eq5}) to eliminate $\psi''$ terms in the stress tensor. Note that this constant does not require $\Omega_0$ to vanish. One explanation for why it was necessary in the framework of the constructed Lagrangian to set $\Omega_0=0$ is the presence of the term proportional to $z = \int \rmd \rho \tan \psi$. At the level of the Lagrangian for $\psi(\rho)$, there is no closed form for $z(\rho)$.

%%%%%%%%%%%%%%%%%%%%%%%%%%%%%%%%%%%%%%%%%%%%%%%%%%%%%%%%%%%%%%%%%%%%%%%%%%%%%%%%
\section{Conclusion}
We have found the first integral of the second-order shape equation of axisymmetric lipid membranes with vanishing $c_0$, $\tilde{\lambda}$, and $\tilde{p}$. In this case, (\ref{eq5}) is invariant under conformal transformations. If we include any one of those parameters, such as $c_0$, it will break the scale invariance we used to find our first integral, likely making it impossible to solve the equation immediately. It is still an interesting challenge to find a first integral for non-vanishing $c_0$, $\tilde{\lambda}$, and $\tilde{p}$.

We obtained the mechanical interpretation of the first integral by using the membrane stress tensor as in \cite{Guven2002}. We find that the first integral $I$ is proportional to $\vs{f}^a \cdot \vs{X} t_a$ when $\Omega_0=0$, where $t_a$ is the surface vector tangent to the meridians of the surface.

\appendix
\section{Detailed calculations for construction of first integrals}
If $\eta$, $\xi$ and $B$ do not explicitly depend on $\psi'$, then (\ref{eq37}) will be:
\begin{eqnarray}
\rho\cos\psi F=0,
\end{eqnarray}
 which means $F=0$. And (\ref{eq38}) is transformed into:
\begin{eqnarray}
  \fl-\rho\cos{\psi}\xi_\psi\psi'^3+\left[\cos{\psi}\left(\xi-\rho\xi_\rho+2\rho\eta_\psi\right)-\rho\sin{\psi}\eta\right]\psi'^2\nonumber\\
  +\left[2\rho\cos{\psi}\eta_\rho-2\left(B_\psi+V\xi_\psi\right)\right]\psi'-2\left(V\xi_\rho+V_\rho\xi+V_\psi\eta+B_\rho\right)=0\label{eq100}
\end{eqnarray}
We can separate the above equation into five independent equations according to the different orders of $\psi'$:
\begin{eqnarray}
  \rho\cos{\psi}\xi_\psi=0,\label{eq101}\\
  \cos{\psi}\left(\xi-\rho\xi_\rho+2\rho\eta_\psi\right)-\rho\sin{\psi}\eta=0,\label{eq102}\\
  \rho\cos{\psi}\eta_\rho-B_\psi-V\xi_\psi=0,\label{eq103}\\
  V\xi_\rho+V_\rho\xi+V_\psi\eta+B_\rho=0.\label{eq104}
\end{eqnarray}
From (\ref{eq101}) and (\ref{eq102}), we obtain:
\begin{eqnarray}
  \eta(\rho,\psi)=\frac{C_1(\rho)}{\sqrt{\cos{\psi}}}\int\sqrt{\cos\psi}\rmd\psi+\frac{1}{\sqrt{\cos{\psi}}}C_2(\rho),\label{eq105}\\
  \xi(\rho)=C_3\rho+2\rho\int{\frac{C_1(\rho)}{\rho}\rmd\rho},\label{eq106}
\end{eqnarray}
Then from (\ref{eq103}), (\ref{eq104}), (\ref{eq33}) and $B_{\psi\rho}=B_{\rho\psi}$, we have:
\begin{eqnarray}
  \fl\frac12\{[35+56\tilde{\lambda}\rho^2-4\left(1+6\tilde{\lambda}\rho^2\right)\cos{2\psi}+\cos{4\psi}\nonumber\\
  +8\rho\sin{\psi}\left(-9c_0-10\Omega_0+5\tilde{p}\rho^2+c_0\cos{2\psi}\right)]C_1(\rho)\nonumber\\
  -32\rho\cos^4{\psi}\left(C_1'+\rho C_1''\right)\}\int\sqrt{\cos\psi}\rmd\psi\nonumber\\
  +\left[8\rho\left(C_3+2\int{\frac{C_1(\rho)}{\rho}\rmd\rho}\right)\right.\nonumber\\
  \left(-c_0-2\Omega_0+3\tilde{p}\rho^2+c_0\cos{2\psi}+4\tilde{\lambda}\rho\sin{\psi}\right)\nonumber\\
  +6\left(-4c_0\rho-8\Omega_0\rho+4\tilde{p}\rho^3+4c_0\rho\cos{2\psi}+5\sin{\psi}\right.\nonumber\\
  \left.\left.+8\tilde{\lambda}\rho^2\sin{\psi}+\sin{3\psi}\right)C_1(\rho)\right]\left(1-\sin^2\psi\right)\sqrt{\sec\psi}\nonumber\\
  +\frac{C_2(\rho)}{2}\left[35+56\tilde{\lambda}\rho^2-4\left(1+6\tilde{\lambda}\rho^2\right)\cos{2\psi}+\cos{4\psi}\right.\nonumber\\
  \left.+8\rho\left(-9c_0-10\Omega_0+5\tilde{p}\rho^2+c_0\cos{2\psi}\right)\sin{\psi}\right]\nonumber\\
  -16\rho\cos^4{\psi}\left(C_2'+\rho C_2''\right)=0.\label{eq107}
\end{eqnarray}
As we have:
\begin{eqnarray}
 \sqrt{\sec\psi}=(1-\sin^2\psi)^{-\frac14},\label{eq119}\\
 \int\sqrt{\cos\psi}\rmd\psi=\int(1-\sin^2\psi)^{-\frac14}\rmd(\sin\psi),\label{wq118}
\end{eqnarray}
 we can expand $\sqrt{\sec\psi}$ and $\int\sqrt{\cos\psi}\rmd\psi$ in terms of $\sin\psi$:
\begin{eqnarray}
 \sqrt{\sec\psi}=\sum_{n=0}^{\infty}\frac{\Gamma(n+\frac14)}{\Gamma(\frac14)\Gamma(n+1)}\sin^{2n}\psi,\label{eq117}\\
 \int\sqrt{\cos\psi}\rmd\psi=G+\sum_{n=0}^{\infty}\frac{\Gamma(n+\frac14)}{(2n+1)\Gamma(\frac14)\Gamma(n+1)}\sin^{2n+1}\psi,\label{eq116}
\end{eqnarray}
where $G$ in (\ref{eq116}) is an integral constant.
Then we can write (\ref{eq107}) in terms of $\sin\psi$:
\begin{eqnarray}
 \sum_{n=0}^4\theta_n\sin^n\psi+\sum_{n=0}^\infty\Theta_n\sin^{2n+5}\psi+\sum_{n=0}^\infty\Lambda_n\sin^{2n+6}\psi=0,\label{eq122}
\end{eqnarray}
where
\begin{eqnarray}
 \fl\theta_0=8\rho\left[\left(-2\Omega_0+3\tilde{p}\rho^2\right)\left(C_3+2\int\frac{C_1(\rho)}{\rho}\rmd\rho\right)+3\left(-2\Omega+\tilde{p}\rho^2\right)C_1(\rho)\right]\nonumber\\
 +16G\left[\left(1+\tilde{\lambda}\rho^2\right)C_1(\rho)-\rho\left(C_1'+\rho C_1''\right)\right]\nonumber\\
 +16\left[\left(1+\tilde{\lambda}\rho^2\right)C_2(\rho)-\rho\left(C_2'+\rho C_2''\right)\right],\label{eq123}\\
 \fl\theta_1=64\tilde{\lambda}\rho^2\int\frac{C_1(\rho)}{\rho}\rmd\rho+\left[64\left(1+\tilde{\lambda}\rho^2\right)+4G\rho\left(-8c_0-10\eta_0+5\tilde{p}\rho^2\right)\right]C_1(\rho)\nonumber\\
 +4\rho\left[8C_3\tilde{\lambda}\rho+\left(-8c_0-10\Omega_0+5\tilde{p}\rho^2\right)C_2(\rho)-4C_1'-4\rho C_1''\right],\label{eq124}\\
 \fl\theta_2=-2\rho\left[8C_3c_0-6C_3\Omega_0+9C_3\tilde{p}\rho^2+2\left(8c_0-6\Omega_0+9\tilde{p}\rho^2\right)\int\frac{C_1(\rho)}{\rho}\rmd\rho\right.\nonumber\\
 +\left(40c_0+2\Omega_0-12G\tilde{\lambda}\rho-\tilde{p}\rho^2\right)C_1(\rho)-16G\left(C_1'+\rho C_1''\right)\nonumber\\
 \left.-16C_2'-4\rho\left(3\tilde{\lambda}C_2(\rho)+4C_2''\right)\right],\label{eq125}\\
 \fl\theta_3=-\frac43\left[18C_3\tilde{\lambda}\rho^2+44C_1(\rho)+\rho\left(36\tilde{\lambda}\rho\int\frac{C_1(\rho)}{\rho}\rmd\rho+6c_0C_2(\rho)\right.\right.\nonumber\\
 \left.\left.+8\tilde{\lambda}\rho C_1(\rho)+6c_0GC_1(\rho)-23C_1'-23\rho C_1''\right)\right],\label{eq126}\\
 \fl\theta_4=\frac{9}{12}C_3\rho\left(16c_0+2\Omega_0-3\tilde{p}\rho^2\right)+4C_2(\rho)+\frac{3}{2}\rho\left(16c_0+2\Omega_0-3\tilde{p}\rho^2\right)\nonumber\\
 \int\frac{C_1(\rho)}{\rho}\rmd\rho+\frac{\rho}{12}\left(304c_0+14\Omega_0-7\tilde{p}\rho^2\right)C_1(\rho)+4GC_1(\rho)\nonumber\\
 -16G\rho\left(C_1'+\rho C_1''\right)-16\rho\left(C_2'+\rho C_2''\right),\label{eq127}\\
 \fl\Theta_n=\frac{1}{\Gamma(\frac14)\Gamma(n+3)}\left\{-48\tilde{\lambda}\Gamma\left(n+\frac54\right)\rho^2\int\frac{C_1(\rho)}{\rho}\rmd\rho+\left[-6C_3\tilde{\lambda}(2n+1)(2n+3)\right.\right.\nonumber\\
 (2n+5)(4n+1)\rho^2-4\left(-135-348n-172n^2+24n^3+16n^4\right)C_1(\rho)\nonumber\\
 -4(n+3)(2n+1)(4n+1)(4n+5)\tilde{\lambda}\rho^2C_1(\rho)-\rho(415+792n+308n^2)\nonumber\\
 \left.\left.(C_1'+\rho C_1'')\right]\frac{\Gamma\left(n+\frac14\right)}{(2n+1)(2n+3)(2n+5)}\right\},\label{eq128}\\
 \fl\Lambda_n=\frac{\Gamma\left(n+\frac54\right)\rho}{2\Gamma\left(\frac14\right)\Gamma(n+4)}\left\{6\left[8c_0(n+3)+(4n+5)\left(2\Omega-3\tilde{p}\rho^2\right)\right]\int\frac{C_1(\rho)}{\rho}\rmd\rho\right.\nonumber\\
 +\left[3C_3(2n+3)(2n+5)\left(8c_0(n+3)+(4n+5)\left(2\Omega_0-3\tilde{p}\rho^2\right)\right)\right.\nonumber\\
 +\left(8c_0\left(n+3\right)\left(85+82n+16n^2\right)+(2n+3)(4n+5)(8n+15)\right.\nonumber\\
 \left.\left.\left.(2\Omega-\tilde{p}\rho^2)\right)C_1(\rho)\right]\frac{1}{(2n+3)(2n+5)}\right\}.\label{eq129}
\end{eqnarray}
And (\ref{eq122}) is equivalent to seven independent equations according to the order of $\sin\psi$:
\begin{eqnarray}
\theta_0=\theta_1=\theta_2=\theta_3=\theta_4=\Theta_n=\Lambda_n=0,
\end{eqnarray}
which leads to the following equations:
\begin{eqnarray}
  \fl\rho\left[\left(-2\Omega_0+3\tilde{p}\rho^2\right)\left(C_3+2\int\frac{C_1(\rho)}{\rho}\rmd\rho\right)+3\left(-2\Omega_0+\tilde{p}\rho^2\right)C_1(\rho)\right]\nonumber\\
  +2G\left[\left(1+\tilde{\lambda}\rho^2\right)C_1(\rho)-\rho\left(C_1'+\rho C_1''\right)\right]\nonumber\\
  +2\left[\left(1+\tilde{\lambda}\rho^2\right)C_2(\rho)-\rho\left(C_2'+\rho C_2''\right)\right]=0,\label{eq108}\\
  \fl16\tilde{\lambda}\rho^2\int\frac{C_1(\rho)}{\rho}\rmd\rho+\left[16\left(1+\tilde{\lambda}\rho^2\right)+G\rho\left(-8c_0-10\eta_0+5\tilde{p}\rho^2\right)\right]C_1(\rho)\nonumber\\
  +8C_3\tilde{\lambda}\rho^2+\rho\left(-8c_0-10\Omega_0+5\tilde{p}\rho^2\right)C_2(\rho)-4\rho C_1'-4\rho^2C_1''=0,\label{eq109}\\
  \fl8C_3c_0-6C_3\Omega_0+9C_3\tilde{p}\rho^2+2\left(8c_0-6\Omega_0+9\tilde{p}\rho^2\right)\int\frac{C_1(\rho)}{\rho}\rmd\rho\nonumber\\
  +\left(40c_0+2\Omega_0-12G\tilde{\lambda}\rho-\tilde{p}\rho^2\right)C_1(\rho)-16G\left(C_1'+\rho C_1''\right)\nonumber\\
  -16C_2'-4\rho\left(3\tilde{\lambda}C_2(\rho)+4C_2''\right)=0,\label{eq110}\\
  \fl18C_3\tilde{\lambda}\rho^2+44C_1(\rho)+\rho\left(36\tilde{\lambda}\rho\int\frac{C_1(\rho)}{\rho}\rmd\rho+6c_0C_2(\rho)\right.\nonumber\\
  \left.+8\tilde{\lambda}\rho C_1(\rho)+6c_0GC_1(\rho)-23C_1'-23\rho C_1''\right)=0,\label{eq111}\\
  \fl9C_3\rho\left(16c_0+2\Omega_0-3\tilde{p}\rho^2\right)+48C_2(\rho)+18\rho\left(16c_0+2\Omega_0-3\tilde{p}\rho^2\right)\nonumber\\
  \int\frac{C_1(\rho)}{\rho}\rmd\rho+\rho\left(304c_0+14\Omega_0-7\tilde{p}\rho^2\right)C_1(\rho)\nonumber\\
  +48GC_1(\rho)-192G\rho\left(C_1'+\rho C_1''\right)-192\rho\left(C_2'+\rho C_2''\right)=0,\label{eq112}\\
  \fl-12\tilde{\lambda}(4n+1)(2n+1)(2n+3)(2n+5)\rho^2\int\frac{C_1(\rho)}{\rho}\rmd\rho\nonumber\\
  -6C_3\tilde{\lambda}(2n+1)(2n+3)(2n+5)(4n+1)\rho^2\nonumber\\
  -4\left[-135-348n-172n^2+24n^3+16n^4\right.\nonumber\\
  \left.+(n+3)(2n+1)(4n+1)(4n+5)\tilde{\lambda}\rho^2\right]C_1(\rho)\nonumber\\
  -\rho(415+792n+308n^2)(C_1'+\rho C_1'')=0,\label{eq113}\\
  \fl6(2n+3)(2n+5)\left[8c_0(n+3)+(4n+5)\left(2\Omega_0-3\tilde{p}\rho^2\right)\right]\int\frac{C_1(\rho)}{\rho}\rmd\rho\nonumber\\
  +3C_3(2n+3)(2n+5)\left[8c_0(n+3)+(4n+5)\left(2\Omega_0-3\tilde{p}\rho^2\right)\right]\nonumber\\
  +\left[8c_0\left(n+3\right)\left(85+82n+16n^2\right)+(2n+3)(4n+5)(8n+15)\right.\nonumber\\
  \left.(2\Omega_0-\tilde{p}\rho^2)\right]C_1(\rho)=0.\label{eq114}
\end{eqnarray}

We note that (\ref{eq113}) and (\ref{eq114}) should valid for any non-negative integer $n$, and expressions of $C_1(\rho)$ and $C_2(\rho)$ do not contain $n$. Then we investigate the coefficient of $n^3$ and $n^4$ in (\ref{eq113}), which should be zero separately, and lead to two equations:
\begin{eqnarray}
 -912\tilde{\lambda}\rho^2\left(C_3+2\int\frac{C_1(\rho)}{\rho}\rmd\rho\right)-32\left(3+20\tilde{\lambda}\rho^2\right)C_1(\rho)=0,\label{eq135}\\
 -192\tilde{\lambda}\rho^2\left(C_3+2\int\frac{C_1(\rho)}{\rho}\rmd\rho\right)-64\left(1+2\tilde{\lambda}\rho^2\right)C_1(\rho)=0.\label{eq136}
\end{eqnarray}
Canceling $C_3+2\int\frac{C_1(\rho)}{\rho}\rmd\rho$ both in (\ref{eq135}) and (\ref{eq136}), we have:
\begin{eqnarray}
 (2\tilde{\lambda}\rho^2-13)C_1(\rho)=0,
\end{eqnarray}
which leaves us no choice but $C_1(\rho)=0$, and hence $C_3\tilde{\lambda}=0$. Then (\ref{eq108})--(\ref{eq112}) will become as:
\begin{eqnarray}
 2\left(1+\tilde{\lambda}\rho^2\right)C_2(\rho)+\rho\left(3C_3\tilde{p}\rho^2-2C_3\Omega_0-2C_2'-2\rho C_2''\right)=0,\label{eq130}\\
 \left(5\tilde{p}\rho^2-8c_0-10\Omega_0\right)C_2(\rho)=0,\label{eq131}\\
 8C_3c_0-6C_3\Omega_0+9C_3\tilde{p}\rho^2-12\tilde{\lambda}\rho C_2(\rho)-16\left(C_2'+\rho C_2''\right)=0,\label{eq132}\\
 c_0C_2(\rho)=0,\label{eq133}\\
 3C_3\rho\left(16c_0+2\Omega_0-3\tilde{p}\rho^2\right)+16C_2(\rho)-64\rho\left(C_2'+\rho C_2''\right)=0.\label{eq134}
\end{eqnarray}
Canceling $\rho(C_2'+\rho C_2'')$ both in (\ref{eq130}) and (\ref{eq132}), we can solve $C_2(\rho)$ as:
\begin{eqnarray}
C_2(\rho)=\frac{C_3\rho\left(8c_0+10\Omega_0-15\tilde{p}\rho^2\right)}{4\left(4+7\tilde{\lambda}\rho^2\right)}.\label{eq137}
\end{eqnarray}
Canceling $\rho(C_2'+\rho C_2'')$ both in (\ref{eq130}) and (\ref{eq134}), we can solve $C_2(\rho)$ as:
\begin{eqnarray}
C_2(\rho)=\frac{C_3\rho\left(48c_0+70\Omega_0-105\tilde{p}\rho^2\right)}{16\left(3+4\tilde{\lambda}\rho^2\right)}.\label{eq138}
\end{eqnarray}
If $C_3=0$, then $C_2(\rho)=0$, which leads to a trivial solution of the first integral. So we should assume $C_3\neq0$, and hence $\tilde{\lambda}=0$, in our discussion. With combining (\ref{eq137}) and (\ref{eq138}), we have:
\begin{eqnarray}
 6c_0+10\Omega_0-15\tilde{p}\rho^2=0,
\end{eqnarray}
which leads to :
\begin{eqnarray}
 \tilde{p}=0,\quad\Omega_0=-\frac35c_0.\label{eq149}
\end{eqnarray}
Then we have:
\begin{eqnarray}
 C_2(\rho)=\frac{c_0C_3}{8}\rho.\label{eq150}
\end{eqnarray}
And because of (\ref{eq133}) we have:
\begin{eqnarray}
 c_0C_2(\rho)=\frac{c_0^2C_3}{8}\rho=0,\label{eq152}
\end{eqnarray}
which leaves:
\begin{eqnarray}
 c_0=\Omega_0=\tilde{p}=C_2(\rho)=0.
\end{eqnarray}
And it is exactly the solution of (\ref{eq107}), we could check this by substituting $C_1(\rho)=C_2(\rho)=c_0=\tilde{\lambda}=\tilde{p}=\Omega_0=0$ into (\ref{eq107}).

So the non-trivial solution for (\ref{eq107}) only exists in the case of vanishing $c_0$, $\tilde{\lambda}$, $\tilde{p}$ and $\Omega_0$:
\begin{eqnarray}
C_1(\rho)=C_2(\rho)=0,\quad C_3\neq0,\label{eq120}
\end{eqnarray}
which leaves the only solution for (\ref{eq100}):
\begin{eqnarray}
B=0,\quad\eta=0,\quad\xi=C_3\rho.\label{eq121}
\end{eqnarray}
\ack
The authors are grateful for the valuable and inspiring discussions with Ping Ao. We thank Markus Deserno for his insightful suggestions. The authors are also grateful for financial support from the National Natural Science Foundation of China (Grant No. 11274046) and the National Science Foundation of the United States (award No. 1515007).

%\bibliographystyle{iopart-num}
%\bibliography{papers}
\end{document}